 \documentclass[10pt,preprint]{aastex}  
 \usepackage{natbib}
\def\ang{\AA}
\def\arcsec{\hbox{$^{\prime\prime}$}}

\def\gapprox{\lower.4ex\hbox{$\;\buildrel >\over{\scriptstyle\sim}\;$}}
\def\lapprox{\lower.4ex\hbox{$\;\buildrel <\over{\scriptstyle\sim}\;$}}

\shortauthors{ASCHWANDEN ET AL 2016}
\shorttitle{Magnetic Energies from IRIS and AIA Data}

\begin{document}

\title{         Tracing the Chromospheric and Coronal Magnetic Field  
		with AIA, IRIS, IBIS, and ROSA Data 	}

\author{        Markus J. Aschwanden$^1$	}

\affil{		$^1)$ Lockheed Martin, 
		Solar and Astrophysics Laboratory, 
                Org. A021S, Bldg.~252, 3251 Hanover St.,
                Palo Alto, CA 94304, USA;
                e-mail: aschwanden@lmsal.com }

\author{        Kevin Reardon$^{2,3,4}$	}

\affil{	        $^2)$ National Solar Observatory,
		3665 Discovery Dr., Boulder, CO, 80300; 
		email: kreardon@nso.edu }

\affil{	  	$^3)$ INAF-Osservatorio Astrofisico di Arcetri,
		Large Enrico Fermi 5, I-50125 Firenze, Italy.}

\affil{		$^4)$ Astrophysics Research Centre, 
		School of Mathematics and Physics, 
		Queen’s University, Belfast BT7 1NN, UK}

\author{        Dave Jess$^{5}$	}

\affil{		$^4)$ Astrophysics Research Centre,
		School of Mathematics and Physics,
		Queen's University Belfast, Belfast BT7 1NN, UK;
		e-mail: d.jess@qub.ac.uk}

\begin{abstract}
The aim of this study is to explore the suitability of chromospheric
images for magnetic modeling of active regions. We use high-resolution
images ($0.1\arcsec$) from the {\sl Interferometric Bidimensional 
Spectrometer (IBIS)} in the Ca II 8542 \ang\ line, the {\sl Rapid 
Oscillations in the Solar Atmosphere (ROSA}) instrument in the
H$\alpha$ 6563 \ang\ line, the {\sl Interface Region Imaging 
Spectrograph (IRIS)} in the 2796 \ang\ line, and compare non-potential
magnetic field models obtained from those chromospheric images
with those obtained from images of the {\sl Atmospheric
Imaging Assembly (AIA)} in coronal (171 \ang , etc.) and
in chromospheric (304 \ang ) wavelengths.
Curvi-linear structures are automatically traced in those images
with the OCCULT-2 code, to which we forward-fitted magnetic field
lines computed with the {\sl Vertical-Current Approximation
Non-Linear Force Free Field (VCA-NLFFF)} code. We find that the
chromospheric images: (1) reveal crisp curvi-linear structures
(fibrils, loop segments, spicules) that are extremely well-suited
for constraining magnetic modeling; (2) that these curvi-linear
structures are field-aligned with the best-fit solution by 
a median misalignment angle of $\approx 4^\circ-7^\circ$; 
(3) the free energy computed from coronal data may underestimate that 
obtained from cromospheric data by a factor of $\approx 2-4$,
(4) the height range of chromospheric features is confined to
$h \lapprox 4000$ km, while coronal features are detected up to
$h \lapprox 35,000$ km; and (5) the plasma-$\beta$ parameter is
$\beta \approx 10^{-5}-10^{-1}$ for all traced features.
We conclude that chromospheric images reveal important
magnetic structures that are complementary to coronal images
and need to be included in comprehensive magnetic field models,
a quest that is not accomodated in standard NLFFF codes.
\end{abstract}


\section{		INTRODUCTION				}

While traditional methods compute the magnetic field in the solar
corona by potential-field extrapolation of the photospheric 
line-of-sight component, or by 
force-free extrapolation of the photospheric 3D vector field,
inconsistencies have been noticed with the observed 
geometry of coronal loops (Sandman et al.~2009; DeRosa et al.~2009), 
which are supposed to trace out the magnetic field in a low 
plasma-$\beta$ corona. Misalignment angles between theoretical NLFFF
solutions and observed loop directions amount to 
$\mu \approx 24^\circ-44^\circ$ (DeRosa et al.~2009). 
Several studies have been carried out to pin down the uncertainties
of nonlinear force-free field (NLFFF) codes, regarding insufficient
field-of-views, the influence of the spatial resolution,
insufficient constraints at the computation box boundaries, and the 
violation of the force-free assumption in the lower chromosphere
(Metcalf et al.~2008; DeRosa et al.~2009, 2015). 
The latter two problems involve the knowledge of the magnetic field
in the chromosphere, which represents a ``missing link'' between the
photosphere and corona. The exploration of this missing link 
as a magnetic interface between photospheric magnetograms and
coronal loops is the main goal of this study, where we
employ chromospheric high-resolution images observed with the
{\sl Interface Region Imaging Spectrograph (IRIS)} (De Pontieu
et al.~2014), the {\sl Interferometric Bidimensional Spectrometer
(IBIS)} (Cavallini 2006; Reardon \& Cavallini 2008; Cauzzi et al.~2008;
Righini et al.~2010; Reardon et al.~2011), and the {\sl Rapid 
Oscillations in the Solar Atmosphere (ROSA}) instrument (Jess et al.~2010). 

What structure and manifestation has the magnetic field in the 
chromosphere? The chromosphere is defined as the cool part of the
solar atmosphere with a temperature range of $T_e \approx 10^4-10^6$ K
that extends from photospheric heights 
$h_{phot}=0$ km (at $\tau_{\lambda=5000 A}$) upward to 
$h_{chrom} \approx 2000$ km according to hydrostatic models (e.g., Vernazza
et al.~1981), or up to $h_{chrom} \approx 5000$ km according to
observed dynamic phenomena (Ewell et al.~1993; Aschwanden et al.~2002).
While the base of the chromosphere is dominated by the thermal 
pressure (with a high plasma-$\beta$, i.e., $\beta = p_{th}/p_{mag} > 1$),
a cross-over to a magnetically dominated transition region
(with a low plasma-$\beta$, i.e., $\beta < 1$)
is believed to take place at a height of $h \approx 1500$ km, which is 
called the {\sl canopy height} (Gabriel 1976). In coronal heights
above, loops are expected to be aligned with the magnetic field. 
Thus we expect to see the footpoints of field-aligned coronal loops 
above the canopy height. In addition we may often see some field-aligned 
dynamic structures, such as fibrils in active regions, spicules 
when observed above the limb, or mottles when observed in the 
Quiet Sun (De Pontieu et al.~2007; Pietarila et al.~2009). 
While there is extensive
literature about the various chromospheric phenomena, we focus
here on their magnetic properties only.  

From measurements of the 3D vector magnetic field using the 
Na I 5896 \ang\ line with the
Stokes Polarimeter at Mees Solar Observatory, it was found that
the magnetic field is not force-free in the photosphere, but
becomes force-free at an altitude of $h \approx 400$ km 
(Metcalf et al.~1995).
However, a lot of chromospheric structures in weak-field regions
are magnetically closed inside the chromosphere (below the canopy height), 
so that there is not always a link between the photospheric network 
and magnetic loops in the corona (Jendersie and Peter 2006). 
The non-potentiality
of the magnetic field in the chromosphere has been measured with
line-of-sight magnetograms at the National Solar Observatory's
Kitt Peak Observatory (NSO/KP) using the 8542 \ang\ line, which is primarily
sensitive to the magnetic field at a height of $h \approx 800$ km
(Harvey et al.~1999; Choudhary et al.~2001). 3D numerical MHD 
simulations with the Bifrost code (Gudiksen et al.~2011) allow us 
to localize the contribution heights of the Ca II 8542 \ang\ line 
in a range of $h \lapprox 1500$ km (Ortiz et al.~2014). 
The 8542 \ang\ line is particularly suited to observe the
fine structure of fibril-like features (Pietarila et al.~2009), 
to measure their geometry and orientation, and to determine
their magnetic field-alignment and non-potentiality
(Jing et al.~2011). The field-alignment of chromospheric fibrils
was tested by comparing CRisp Imaging Spectro-Polarimeter data
(CRISP, Scharmer 2006) with Spectro-Polarimeter for INfrared and 
Optical Regions data (SPINOR, Socas-Navarro et al.~2006), and it
was found that fibrils are often oriented along the magnetic field,
but not always (de la Cruz Rodriguez and Socas-Navarro 2011).
Also in MHD simulations it was found that some modeled fibrils
are not field-aligned in that simulated volume (Leenaarts et al.~2015).
IBIS observations from the Dunn Solar Telescope (DST) in New Mexico
found that fibrils are aligned with the magnetic field with an
uncertainty of $\mu \lapprox 10^\circ$ (Schad et al.~2013).

In this study we conduct nonpotential field modeling of
chromospheric structures and coronal loops with a particular
NLFFF code that is based on a {\sl vertical-current approximation
(VCA)}, and we perform automated tracing of coronal loops and 
chromospheric curvi-linear features. References to technical 
descriptions of the VCA-NLFFF code are provided in Section 2. 
Data analysis of {\sl Atmospheric Imager Assembly (AIA)} 
(Lemen et al.~2012) and {\sl Helioseismic Magnetic Imager (HMI)} 
(Scherrer et al.~2012) data from the {\sl Solar Dynamics 
Observatory (SDO)} (Pesnell et al.~2011), from the 
{\sl Interface Region Imaging Spectrograph (IRIS)}
(De Pontieu et al.~2014), from the {\sl Interferometric 
Bidimensional Spectrometer (IBIS)} (Cavallini 2006), and
from the {\sl Rapid Oscillations in the Solar Atmosphere (ROSA)}
(Jess et al.~2010) instrument are presented in Section 3, 
and discussion and conclusions are offered in Sections 4 and 5. 

\section{		METHOD			}

Traditional NLFFF
codes use the 3D vector field ${\bf B}(x,y)=[B_x(x,y), B_y(x,y),
B_z(x,y)]$ from a vector magnetograph instrument as input for
the photospheric boundary (in the $(x,y)$-plane), and use a
height-extrapolation scheme to compute magnetic field lines that
match the boundary condition, the divergence-freeness, and the
force-freeness conditions. 
Examples and comparisons of such recent NLFFF codes 
are given in Metcalf et al.~(2008) and DeRosa et al.~(2009, 2015),
which includes the optimization method (Wheatland et al.~2000;
Wiegelmann and Inhester 2010), the magneto-frictional method
(Valori et al.~2007, 2010), the Grad-Rubin method (Wheatland
2007; Amari et al. 2006), and others. 

Here we use an alternative method which is called the {\sl Vertical
Current Approximation Nonlinear Force-Free Field (VCA-NLFFF)} method.
The theoretical model assumes a variable amount of vertical currents
associated with each magnetic field concentration, which introduces 
a helical twist about the vertical axis (Fig~1).
A detailed description and performance tests of measuring the
non-potential magnetic energy is given in a recent study
(Aschwanden 2016). This code uses only the line-of-sight magnetogram 
$B_z(x,y)$ to constrain the potential field, and does forward-fit an 
analytical approximation of a vertical-current solution to automatically 
traced coronal or chromospheric loop coordinates $[x(s),y(s)]$ to 
determine the nonlinear force-free $\alpha$-parameters for a number of
unipolar (subphotospheric) magnetic sources. 
The chief advantages of the VCA-NLFFF code over traditional NLFFF
codes are the circumvention of the unrealistic assumption of a
force-free photosphere in the magnetic field extrapolation method,
the capability to minimize the misalignment angles between observed
coronal loops (or chromospheric fibril structures) and theoretical
model field lines, as well as computational speed.
The theory of the vertical-current approximation is originally derived 
in Aschwanden (2013a), while the numerical VCA-NLFFF 
code has been continuously developed and improved
in a number of previous studies (Aschwanden and Sandman 2010;
Sandman and Aschwanden 2011; Aschwanden et al.~2008, 2012, 2013, 2014a, 2014b;
Aschwanden 2010, 2013b, 2013c, 2015; Aschwanden and Malanushenko 2013).

A related forward-fitting code, using a quasi-Grad-Rubin method
to match a NLFFF solution to observed coronal loops,
was pioneered by Malanushenko et al.~(2009, 2011, 2012, 2014) also. 

\section{	OBSERVATIONS AND DATA ANALYSIS				}

Since the motivation of this study is the exploration of chromospheric
structures regarding their suitability for magnetic field modeling,
we choose observations with the highest available spatial resolution
that show crisp chromospheric structures. We found such suitable data
from the IBIS instrument in the Ca II 8542 line, and from both the IBIS
and ROSA instrument in the 6563 H$\alpha$ line, both having a spatial 
resolution of $\approx 0.1\arcsec$. We analyze such data from three 
different observing runs (Table 2) at the NSO/SP Dunn Solar Telescope (DST) 
(2010 Aug 3; 2014 Aug 24, 2014 Aug 30), and complement these observations 
with IRIS, AIA/SDO, and HMI/SDO data.

\subsection{	Observations of 2010 Aug 3 		}

The {\sl Interferometric Bidimensional Spectrometer (IBIS)}, a ground-based
dual Fabry-Perot interferometer, records images in the wavelength range of 
5400-8600 \ang\ (Cavallini 2006; Reardon and Cavallini 2008; Righini et al.~2010;
Reardon et al.~2011), is installed on the Dunn Solar Telescope at the NSO/SP
facility, has a field-of-view of $95\arcsec$, a mosaic pixel size of $0.0976\arcsec$,
and acquired images for active region NOAA 11092 on 2010 Aug 3, 15:03-15:43 UT. 
We analyze a $240\arcsec \times 240\arcsec$ mosaic image taken in the 
wavelength of Ca II 8542 \ang\ and H-alpha 6563 \ang\ (described also 
in Reardon et al.~2011; Jing et al.~2011), but we limit the field-of-view 
to a square with a size of FOV=0.10 $R_{\odot}$ centered on the primary 
sunspot of NOAA 11092, covering the ranges of $x=[-0.0159,0.0841]$ $R_{\odot}$ in EW 
direction and $y=[0.0608, 0.1608]$ $R_{\odot}$ in NS direction, centered 
at a heliographic position of N12W02, almost near the center of the solar disk. 
Since IBIS images cover a limited portion of the solar disk only, the coalignment
with full-disk images was carried out with an AIA 1600 \ang\ image at the
mid-time of the IBIS image acquisition time interval, at 15:23:00 UT.
A rendering of the analyzed subimage of IBIS 8542 \ang\ is shown in Fig.~2, 
where the umbra of the sunspot appears dark, surrounded by a wrath of 
fibrils that spiral in curved trajectories from the penumbra away from 
the sunspot, similar to the non-potential field model (based on vertical
currents) shown in Fig.~1 (bottom left panel), though with opposite chirality.

A highpass-filtered version of the original IBIS 8542 image (Fig.~2) is
shown in Fig.~3 (top panel), which is produced by subtracting a 
finestructure image (the original image smoothed with a boxcar of 
$nsm_1=3$ pixels) from a lowpass-filtered image (the original image 
smoothed with a boxcar of $nsm_2=nsm_1+2=5$ pixels). This filter enhances 
structures with a width of $w \approx 3$ pixel $\approx  0.3\arcsec 
\approx 200$ km. We perform an automated loop tracing run with the 
OCCULT-2 code, setting the maximum number of analyzed structures to
$n_{struc}=10,000$, while all other control parameters are set to
standard values (Table 1).
This run identifies a total of 1193 curvi-linear structures (Fig.~3, bottom
panel). Note the absence of curvi-linear fine structure in the sunspot umbra,
while fibrils or loop segments are filling most of the penumbral area and
surrounding outskirts of the active region. The richness of curvi-linear
structures makes the image highly suitable for magnetic field modeling
and quantitative tests of field-alignment. For a comparison of the
sensitivity and efficiency of the automated feature detection algorithm OCCULT-2,
see also Fig.~2 in Jing et al.~(2011), where the same image has been
processed with a union-finding segmentation algorithm (Sedgewick 2002).

In order to complement the chromospheric image of IBIS with coronal images,
we include AIA/SDO images, at coronal wavelengths of 94, 131, 171, 193, 211, 335
\ang , as well as 304 \ang\ at cooler chromospheric temperatures (see
temperature ranges in Table 2).  AIA subimages
with the same FOV as the IBIS image (Figs.~2-3) are shown in Fig.~4
(left column), with highpass-filtered versions (Fig.~4, middle column),
and automated loop tracings (Fig.~4, right column). While the
original images show fans that contain bundles of loops rooted in the
sunspot, the highpass-filtered images reveal loops with narrower widths,
which are detected and localized with the OCCULT-2 code. It appears that
the radially extending coronal loops seen in AIA have a similar orientation
as the chromospheric structures seen in the IBIS image. However, since IBIS
has a six time higher spatial resolution than AIA on a 1-D linear scale
(corresponding to 36 times more details in a 2-D area), the detected loop 
features are much less numerous in the AIA images ($\approx 15-40$ per 
wavelength in Fig.~4) than in the IBIS image ($n_{IBIS} \approx 1200$ in Fig.~3).
This is mostly a consequence of the much higher spatial resolution of IBIS. 
The 36 times larger amount of image area information in IBIS let us expect 
about $n_{AIA,pred} \approx 1200/36=33$ loop structures per AIA image, which 
is indeed the case, according to the counts of $n_{AIA}\approx 15-40$ in Fig.~4. 
Unfortunately, no contemporaneous IRIS images were available at this time.

\subsection{	Observations of 2014 Aug 24 		}

A second data set that provides ROSA H$\alpha$ images at 6563 \ang\
was obtained during a campaign at the NSO/SP Dunn Solar Telescope
on 2014 Aug 24, 13:55-15:56 UT. We select an image obtained at 14:06:38 UT,
which is also shown in Fig.~2 of Jess et al.~(2015). We select a subimage
with a square field-of-view of size FOV = 0.10 $R_{\odot}$ that centers on the primary 
sunspot of NOAA 12146, covering the ranges of $x=[0.3674,0.4674]$ $R_{\odot}$ in EW 
direction and $y=[-0.0112, 0.0888]$ $R_{\odot}$ in NS direction, which is centered 
at a heliographic position of N09W25, being about 0.4 $R_{\odot}$ away from
disk center. Unfortunately, no cotemporaneous IRIS images are available at this 
time either.

\subsection{	Observations of 2014 Aug 30 		}

A third data set of ROSA H$\alpha$ images during the same week of the the
observing campaign was obtained on 2014 Aug 30, 14:37-17:46 UT.
We select a subimage with a square field-of-view of size FOV = 0.10 $R_{\odot}$ 
that centers on the primary sunspot of NOAA 12149, covering the ranges of 
$x=[0.6295,0.7295]$ $R_{\odot}$ in EW 
direction and $y=[0.0375, 0.1375]$ $R_{\odot}$ in NS direction, which is centered 
at a heliographic position of N12W44, being about 0.7 $R_{\odot}$ away from
disk center. Such large distances from disk center may introduce larger errors
in the deconvolution of magnetograms into unipolar magnetic charges (although
our deconvolution technique takes the 3-D effects into account), as well as
be less suitable for automated tracing of curvi-linear features due to the
confusion caused by a larger number of superimposed structures and a higher
degree of foreshortening. Nevertheless, we include these less than ideal
data in our analysis in order to test the accuracy of magnetic energy
measurements with our VCA-NLFFF code. 

In addition, IRIS (as well as Hinode SOT/SP) was pointing to the same
active region at this time, so that we have an independent comparison
of chromospheric images with ROSA. From IRIS we choose the Mg II h/k line 
at 2796 \ang\ for our analysis, which is sensitive to a chromospheric
temperature range of $log(T)[k])=3.7-4.2$ (or 5000-16,000 K). 

\subsection{	Results of Magnetic Modeling  			}

We perform magnetic field modeling of the 2010 Aug 3, 15:23 UT data in four
different wavelength groups, in the coronal lines of 
AIA/SDO (94, 131, 171, 193, 211, 335 \ang ) (Fig.~5), 
in the AIA/SDO wavelength of He II 304 \ang\ (Fig.~6), 
in the chromospheric IBIS 8542 \ang\ image (Fig.~7), and in the 
IBIS H$\alpha$ 6563 \ang\ line (Fig.~8). The automatically traced loop-like
structures are visualized with yellow curves in Figs.~5-8, while the
best-fit magnetic field lines obtained with the VCA-NLFFF code are shown
in red color. The various input parameters of the VCA-NLFFF code are listed
in Table 1.  We summarize the key results of our magnetic modeling runs 
in Table 3, which includes the number of (automatically detected)
field-aligned loop structures $n_{det}$, the number of loops auto-selected
for fitting $n_{loop}$ (by imposing a progressive elimination of structures
with large misalignment angles during the forward-fitting procedure), 
the median misalignment angles ($\mu_2, \mu_3$), the total potential magnetic
field energies ($E_P$), and the ratios of the non-potential to the potential
magnetic energy, which quantifies the free energy ratio ($q_{free}=E_{NP}/E_P-1$).
In order to estimate the uncertainty of the results,
each of the 11 forward-fits listed in Table 3 has been repeated five times 
with different input parameters ($l_{min}=4, 5$; $n_{mag}=30, 50$, $n_{sm1}=1, 3$, 
and $\mu_0=20^\circ, 30^\circ$), and the mean and standard deviations are listed
in Table 3.

For coronal wavelengths (AIA 94, 131, 171, 193, 211, 335 \ang ) 
we find a large number of 
field-aligned structures, in the range of $n_{det} \approx 200$ (Table 3), 
which is expected since coronal loops are supposed to follow the magnetic field. 
A new result is that we find a large number of field-aligned structures (fibrils) 
in the IBIS Ca II 8542 \ang\ line
($n_{det} \approx 700$; Fig.~7), in the IBIS H$\alpha$ 6563 \ang\ line 
($n_{det} \approx 700$; Fig.~8),
and in the IRIS Mg II 2796 \ang\ line ($n_{det} \approx 700$, Table 3). 
The accuracy of the field alignment is measured with the 2D misalignment
angle $\mu_2$, amounting to $\mu_2 \approx 4^\circ-5^\circ$ for IBIS (Fig.~7 and 8). 
The number of field-aligned loops is lowest for AIA 304 \ang\ images, in 
the range of $n_{det} \approx 50$ (Table 3). 
Moreover, Fig.~6 (top panel) shows that the altitudes of the reconstructed
field-aligned loops are in coronal rather than chromospheric heights.
The small amount of detected structures indicates that most of the structures 
seen in the chromospheric He II line (with a temperature of $log(T[K])=4.7$ 
or $T \approx 50,000$ K) are not field-aligned, curvi-linear, or loop-like. 
Alternatively, a smaller number of detected structures in the He II 304 \ang\
line could be attributed to a higher opacity or the line formation process.

Since the line-of-sight coordinate $z$ cannot be measured, we can observe
the 2-D misalignment angles $\mu_2$ only, which are expected to be in the 
range of $\mu_2 = \sqrt{2/3} \mu_3 \approx 0.8 \mu_3$ for isotropic
errors. The measured median values have a ratio of $\mu_2/\mu_3 \approx 0.6$
(Table 3), where $\mu_2$ is measured from the 2D projected loop angles,
while $\mu_3$ is inferred from the 3D fits of the model field lines.
This indicates that the altitude errors in the magnetic field
models are larger than the horizontal errors. Nevertheless, it is gratifying
to see that through the use of high-resolution images 
the errors of the theoretical (vertical-current) magnetic field
model combined with the errors of automated curvi-linear tracing could be
beaten down to an unprecedented low value of $\mu_2 = 4^\circ-5^\circ$,
compared with the discrepancy of $\approx 24^\circ-44^\circ$ found between 
traditional NLFFF models and observed loops (Sandman et al.~2009; 
DeRosa et al.~2009). There is no other NLFFF model known that matches the 
loop directions with such a high degree of accuracy. 

The most interesting parameter is the free energy ratio in active regions.
The most accurate measurements of the free energy were achieved for the first
case (NOAA 11092 observed on 2010 Aug 3), where the coronal AIA data yield a 
free energy ratio of $q_{free}=3\%\pm1\%$, while the IBIS Ca II 8542
\ang\ and H-alpha 6563 \ang\ data yield 
a significantly larger value of $q_{free}=13\%\pm4\%$, 
and $q_{free}=11\%\pm1\%$, respectively.
It is remarkable that the free energies measured with Ca II 8542 \ang\ 
and H$\alpha$ 6563 \ang\ agree within 20\%, which indicates that the
traced structures are nearly identical, radiate at a similar temperature,
and originate in similar chromospheric heights. For a direct comparison
of field-aligned structures see Fig.~9. 
The traced structures appear to be preferentially rooted in the penumbra,
are absent in the umbra, and are sparse outside the penumbra.
We performed also measurements with AIA 304 \ang\ data and obtained a 
value of $q_{free}=6\%\pm3\%$ that is intermediate to the relatively higher
free energy in the chromosphere and the relatively lower free energy
in the corona. Since the coronal data lead to a significantly lower 
amount of free energy than the chromospheric data, we conclude that 
the two datasets are susceptible to different subsets of magnetic 
field lines, which are governed by different degrees of non-potentiality. 
Chromospheric fibrils and footpoints of loops apparently reveal a 
higher degree of helical twist than coronal loops.

The second case (NOAA 12146 observed on 2014 Aug 24) exhibits a free energy ratio
of $q_{free}=26\%\pm1\%$ (Table 3) from the chromospheric ROSA H-alpha image, 
while the values from the coronal AIA data are lower also, 
but somewhat less reliable,
given the relatively large misalignment angle of $\mu_2 \gapprox 6^\circ$ that
indicates insufficient convergence, either due to a lack of suitable loops,
due to the complexity of nested loops, or due to confusing projection effects 
at a heliographic longitude of W25. 

The third case (NOAA 12149 observed on 2014 Aug 30) shows a free energy ratio of
of $q_{free}=26\%\pm1\%$ from the chromospheric IRIS 2796 \ang\ data, and 
of $q_{free}=17\%\pm4\%$ from the chromospheric ROSA H$\alpha$ data,
while the coronal AIA data show again a lower value of
$q_{free}=10\%\pm5\%$ (Table 3).

In summary, we find that the coronal data tend to underestimate the total
free energy of an active region in all three cases, by a typical factor
of $\approx 2-4$.  The highest accuracy or smallest misalignment
angle is achieved for heliographic positions near disk center 
($\mu_2 \approx 4^\circ$) , while
heliographic positions all the way to a distance of $0.7 R_{\odot}$ become
gradually less accurate ($\mu_2 \approx 7^\circ$). 
This may indicate that the 3D geometry appears to 
be more nested and confused near the limb, imposing a bigger challenge
for accurate 3D reconstruction. In addition, the contrast of the
structures in the H-alpha images are higher in the IBIS images than
in the ROSA images due to the narrower passband of IBIS.
Nevertheless, since we achieve a
similar accuracy for the best-fit median misalignment angle ($\mu_2$, 
$\mu_3$) for both chromospheric structures and coronal loops we can conclude 
that chromospheric fibrils (or footpoints of coronal loop structures) 
are generally field-aligned, within an accuracy of 
$\mu_2 \approx 4^\circ-7^\circ$.

\subsection{	Altitudes of Chromospheric Tracers			}

Since the best-fit solution of our VCA-NLFFF code provides the 
line-of-sight coordinate $z$ for every loop tracing position $(x,y)$,
we can directly determine the altitude range $h=r-1=\sqrt{x^2+y^2+z^2}-1$
for each traced loop segment. This allows us to measure an altitude
distribution of all traced loop segments in a given wavelength range,
which corresponds to the contribution function (as a function of
height) in each wavelength. An approximate range of the altitude distribution
of coronal and chromospheric tracers can already be seen in the side
views shown in Figs.~5-8. Coronal loop tracings cover most of the
altitude range of $h \lapprox 0.05 R_{\odot} = 35,000$ km (Fig.~5 and 6 top panels), 
while chromospheric tracers extend only over a chromospheric height 
range of $h \lapprox 0.005 R_{\odot} = 3500$ km (Fig.~7 and 8, top panels),
although we used an identical computation box with a height range
of $h_{max}=0.05 R_{\odot} = 35,000$ km in fitting each data set. 

We plot histograms of the altitudes $h$ of all automatically traced 
loop segments (or chromospheric features) for each instrument or 
wavelength separately, where each segment with a length of
$\Delta s = 0.002 R_{\odot}$ is counted as an individual element, 
which is shown in Fig.~10.
Interestingly, both the hot coronal structures detected
with AIA mostly in 171 and 193 \ang , as well as the cooler structures
detected with AIA at 304 \ang\ are found to have coronal altitudes.
It appears that the He II 304 \ang\ line traces a lot of cool plasma
in coronal structures, either from static cool loops, or from cooling
loops that produce ''coronal rain'', or cool plasma of quiescent 
prominence and filaments, rather than cool plasma in chromospheric heights.
On the other side, the IBIS 8542 \ang , the ROSA H$\alpha$ 6563 \ang ,
and the IRIS Mg II h/k line detect cool plasma only in the chromosphere
or lower transition region with an upper limit of $h \lapprox 4000$ km.
It appears that our VCA-NLFFF magnetic modeling method is equally suited 
to measure altitudes of chromospheric and coronal features
as stereoscopic and tomographic methods (e.g., see review of 
Aschwanden 2011).

\subsection{	Plasma-$\beta$ Parameter of Chromospheric Tracers	}

Our VCA-NLFFF model provides the magnetic field strength $B$
in each automatically traced curvi-linear feature, be it a coronal
loop or a chromospheric fibril. Assuming pressure balance in the chromosphere,
we can then use a hydrostatic chromospheric density $h_e(h)$ and
temperature model $T_e(h)$ in order to determine the plasma-$\beta$
parameter as a function of height $h$,
\begin{equation}
	\beta(h)={p_{th}(h) \over p_{mag}(h)}
	=6.94 \times 10^{-15} n_e(h) T_e(h) B(h)^{-2} \ .
\end{equation}
For a coronal density of $n_e \approx 10^9$ cm$^{-3}$ and a coronal
temperature of $T_e \approx 10^6$ K we thus expect for magnetic
field strength in the range of $B \approx 100-1000$ G a fairly low
value of $\beta \approx 10^{-3}-10^{-5}$, which explains the perfect
plasma confinement in the corona, that holds for most of the
upper chromosphere also (Gary 2001).

Using the VAL-C model (Vernazza et al.~1981; Fontenla et al. 1990) 
for the chromosphere in a range of $h\approx 0-2000$ km and the 
coronal canopy model of Gabriel (1976) in the height range of
$h=2000-100,000$ km, we calculate the plasma-$\beta$ parameter
$\beta (h)$ as a function of the height $h$, based on our magnetic
field solutions $B(h)$ for each automatically traced loop or
chromospheric feature. We show the results in Fig.~11 and see that the 
magnetic field strength in all traced features varies in the
range of $B \approx 100-1000$ G (Fig.~11, left panels). Of course,
there are areas with lower field strengths outside of the analyzed
sunspots and active regions, but it appears that all (automatically)
traced loop structures and fibrils are anchored in strong-field regions, 
while we find virtually no loop rooted in areas with 
$B \lapprox 50$ G, a finding that is also consistent with the
$\beta=1$ contour outside the penumbral region in another case
(Jess et al.~2013, Fig.~1 therein). This
result in itself may have important consequences for coronal heating
models. Using then these field strengths $B(h)$ and combining with
the temperature $T_e(h)$ and density models $n_e(h)$ we obtain then
with Eq.~(1) the variation of the plasma-$\beta$ parameter with
height, which is shown in the Fig.~11 (right-hand panels). Apparently
the values of the plasma-$\beta$ parameter are below unity for all
traced loop structures, spreading over a height range of
$h \approx 100-20,000$ km. This applies not only to coronal loops,
but also to all chromospheric structures (fibrils) observed here
with IBIS, ROSA, and IRIS. The observed preference for low 
plasma-$\beta$ values implies locations inside magnetic field
concentrations, which does not exclude higher plasma-$\beta$ values
outside the observed structures.

\section{	DISCUSSION  				}

\subsection{	Chromosphere Suitability for Magnetic Modeling  	}

The major motivation for this study is the question whether
chromospheric data are useful for magnetic field modeling.
From previous modeling we know that coronal data that show
the geometry of loops are highly useful for NLFFF modeling.
DeRosa et al.~(2015) conclude in their latest NLFFF modeling
comparison: ``We continue to recommend verifying agreement
between the modeled field lines and corresponding coronal 
loop images before any NLFFF model is used in a scientific
setting''. On the other side we are painfully aware of the
limitations of photospheric data, regarding the actual
non-forcefreeness that violates the forcefree assumption 
used in NLFFF extrapolations from the photospheric boundary
(DeRosa et al.~2009). So, what about the chromosphere,
which is situated in the interface between the photosphere
and the corona? In particular to cicrumvent the forcefreeness
dilemma, the question was posed by Wiegelmann: ``Can we improve the 
preprocessing of photospheric magnetograms by the inclusion 
of chromospheric observations?'' (Wiegelmann et al.~2008). 
This idea was tested with a model image of H$\alpha$ fibrils 
and was found to improve the NLFFF solutions in a model 
chromosphere (Wiegelmann et al.~2008; Metcalf et al.~2008).

Since the solar atmosphere was found to be
force-free above $h \approx 400$ km (Metcalf et al.~1995),
the force-freeness is not violated in the 
chromosphere. Since the altitudes of
chromospheric structures observed with IBIS, IRIS, and ROSA
is consistent with a height range of $h \approx 400-4000$ km
as measured with our VCA-NLFFF code (Fig.~11), and moreover
the plasma-$\beta$ parameter was found to be less than unity
in the entire height range of $h\approx 100-35,000$ km,
chromospheric features should be field-aligned as well as
not violate the force-freeness condition, a finding that
is consistent with another sunspot study (Jess et al.~2013;
Fig.~1 therein). Another confirmation
that chromospheric features are field-aligned is corroborated
with our finding that the automatically traced chromospheric
features (fibrils and footpoints of loops) match the best-fit 
VCA-NLFFF solutions as close as coronal data do, for instance
the 2-D misalignment for chromospheric data is about
$\mu_2 \gapprox 4^\circ$ (Table 3).
This match is even more accurate than the field-alignment of
super-penumbral fibrils measured with IBIS 8542 \ang\ 
determined with a Hanle and Zeeman modeling code, which was
found to be $\mu_2 \approx 10^\circ$ (Schad et al.~2013),
or with SPINOR/CRISP observations, which show a median
misalignment of $\mu \approx 20^\circ$ (de la Cruz Rodriguez and 
Socas-Navarro 2011; their Table 2). 

Thus, at a first glance,
we can answer the question of the suitability of chromospheric
data for magnetic modeling clearly in an affirmative way,
based on their force-free nature, their low plasma-$\beta$
parameter, and the small misalignment angle range found with the
best-fit VCA-NLFFF solutions. However, we have to add a caveat
that our measurements mostly apply to strong-field regions
(although field-aligned fibrils are seen in some plage regions also),
while the appearance of field-aligned fibrils may be 
more chaotic and less force-free in weak-field regions such as
in the Quiet Sun, rendering chromospheric data possibly
less useful for magnetic modeling there (Wiegelmann et al.~2015).

\subsection{	Chromospheric versus Coronal Free Energy 	}

Ideally, the volume-integrated free energy in an active region
should match for chromospheric and coronal data, if a perfect
coronal magnetic field model is available. 
A good correspondence of the free energy between chromospheric data 
(IRIS, AIA 304, 1600 \ang ) and coronal data (AIA 94, 131, 171, 
193, 211, 335 \ang ) was indeed found during a solar flare, 
where the time evolution of the free energy agreed in magnitude 
as well as exhibited a synchronized increase and decrease in
coronal and chromospheric data (Aschwanden 2015), based on an
earlier version of the VCA-NLFFF code used here.

In the present study we notice a substantial disagreement between
the free energy of an active region computed from chromospheric 
versus coronal tracers, where the coronal data reveal a trend to
underestimate the free energy. Obviously,
the two data sets with automatically traced structures do not
constitute representative samples, but rather may contain
mutually exclusive fractions, one being sampled in the chromosphere
at altitudes of $\lapprox 4000$ km, while the other is sampled
in the corona up to $h \lapprox 35,000$ km. The two samples
appear to have a different degree of magnetic twist, and
therefore yield a different amount of free energy. The
chromosphere seems to contain stronger twisted loops than
the corona, which results into a higher free energy. 
In future applications we might combine the chromospheric and
coronal data in a single forward-fitting procedure, so that a
more complete and more representative sample of field-aligned
structures is available. For the time being we have to conclude
that the coronal data can underestimate the free energy (as
determined with the VCA-NLFFF code), while the chromospheric 
data appear to sample the magnetic field more comprehensively.

\subsection{	Chromospheric Height Contribution Function 	}

Understanding the 3-D geometry of chromospheric structures (fibrils,
loops, moss (Berger et al. 1999, etc.) in the context of 3-D magnetic field modeling
requires the height contribution function of each
observed wavelength. The atmospheric height of maximum emissivity 
or absorption of an optically-thick line strongly varies from the line core to 
the line wings. Measurements of the chromospheric magnetic field
at several wavelengths within the Na I line, by computing the
net Lorentz force as a function of wavelength (using the equations
of Molodenskii 1969) demonstrated the height dependence of the force,
since the wings of the Na I line are formed deeper in the
atmosphere than the core (Metcalf et al.~1995). Combining this 
magnetic height dependence $B(h)$ with an atmospheric model
of the density and temperature (VLA-F model), the contribution 
function of the Na I line was found to extend over a height range
of $h \approx 100-1000$ km (Fig.~6 of Metcalf et al.~1995). 
For the Ca II 8542 \ang\ line, the best correlation between 
a chromospheric magnetogram and a magnetic potential field model 
was found at 800 km, which represents the peak of the line formation 
(Choudhary et al.~2001; Leenaarts et al.~2010). 

The Ca II 8542 \ang\ contribution 
function was also calculated with the FAL-C model 
(Fontenla et al.~1993) and with hydrodynamic simulations
by Carlsson and Stein (1997), yielding a range of
$h \approx 300-1000$ km for the bulk of the Ca II emission
(Fig.1 in Pietarila et al.~2009), and $h \lapprox 1300$ km
(Fig.~5 in Cauzzi et al.~2008). Comparing these height ranges
with our 3-D magnetic field reconstruction (with VCA-NLFFF),
we find a somewhat larger height range of $h \lapprox 4000$ km
for the total of all automatically traced curvi-linear features
in IBIS, ROSA, and IRIS data, which may be explained by 
dynamic phenomena that populate the upper chromosphere of
$h \approx 2000-4000$ km, in excess of the static chromospheric
models that span over a lower height of the chromosphere
($h \lapprox 2000$ km for VAL and FAL models). However, the
extended chromosphere is filled with ubiquitous spicules
and fibrils up to $h \lapprox 5000$ km, according to radio,
hard X-ray, and and UV observations (Ewell et al.~1993;
Aschwanden et al.~2002).  H$\alpha$ filtergrams show fibrils up 
to heights of $h \approx 3000-4000$ km (Harvey et al. 1999; 
Choudhary et al.~2001). In any case, 3-D magnetic field
modeling provides height information that is complementary
to the contribution functions determined with hydrodynamic 
models. On the other side, (M)HD models and (N)LTE radiative
transfer calculations are based on the nature of the input
model atmosphere, and hence provides information on the
contribution functions only in a statistical sense.

\section{		CONCLUSIONS 		}

We explored the suitability of chromospheric images for magnetic
modeling, using high-resolution images from IBIS and ROSA in the Ca II 
line at 8542 \ang , and in the H$\alpha$ 6563 \ang\ line,
and from the IRIS Mg II line, and compared the results with coronal 
images taken with AIA/SDO. Our investigation made use of a novel magnetic
field calculation method by fitting field lines that are parameterized 
by an approximative solution of the nonlinear force-free field based on
a vertical-current approximation (the so-called VCA-NLFFF code). 
The field lines of the theoretical model are forward-fitted to
curvi-linear (loop-like) structures that are automatically detected
with the OCCULT-2 code in chromospheric or coronal images. 
We applied this VCA-NLFFF code to observations of active regions 
observed on three days (2010 Aug 3; 2014 Aug 24 and 30) during
IBIS and ROSA campaigns. Since the previous application of the 
VCA-NLFFF code to coronal images from TRACE and AIA has proven the
suitability of coronal images for magnetic field reconstruction
methods, we aim to test here the same application to chromospheric
images. Our conclusions from this first exploration of chromospheric
data in this context are as follows:

\begin{enumerate}

\item{\underbar{The suitability of chromospheric images for magnetic modeling:}
Chromospheric images with high spatial resolution ($0.1\arcsec$), 
especially from the core of the Ca II line (8542 \ang ) and from the core 
of the H$\alpha$ line (6563 \ang ), reveal a wealth of crisp curvi-linear
structures (fibrils and loop segments) and thus are extremely well-suited 
to constrain magnetic field solutions. The suitability has been measured
in terms of the 2D and 3D misalignment angle between the magnetic field
model and the observed field directions, as obtained from automated
detection of curvi-linear structures. For the best case of a sunspot
region near the center of the solar disk, a misalignment angle of
$\mu_2 \approx 4^\circ$ was achieved for a chromospheric image 
(from IBIS), while AIA data yield a similar value of $\mu_2 \approx 5^\circ$  
Other cases at a larger distance to disk center (at
W25 and W44) yielded less accurate values ($\mu_2 \approx 5^\circ-7^\circ$. 
This work seems to indicate
that the orientation of the fibrils seen in the chromospheric images
seem to be consistent with a realistic model of the magnetic field
in the chromosphere, which is in contrast to previous work that showed
large misalignments between fibrils and magnetic field models
(de la Cruz Rodriguez Socas-Navarro 2011; Schad et al.~2013.}

\item{\underbar{The free (magnetic) energy} in an active region, 
i.e., the difference between 
the non-potential and potential magnetic energy, is an important upper 
limit for the global energy that can be dissipated in a flare or CME.
For the most accurate of our three cases we found a free energy ratio
of $q_{free}=13\%\pm4\%$ and $q_{free}=11\%\pm1\%$.
These values correspond to the magnetic energy integrated over a 
computation box with a field-of-view of $0.1 R_{\odot}$,
where the total potential energy is $E_P = 5.71 \times 10^{32}$ erg.
These values are significantly higher than what we find from coronal
tracers, i.e., $q_{free}=3\%\pm1\%$, using combined AIA images at 
the wavelengths of 94-335 \ang . 
Thus, the chromospheric data probe a higher degree of non-potential,
helically twisted structures (fibrils and loop footpoint segments) at
chromospheric temperatures, while the loops imaged in coronal temperatures 
appear to be less twisted and underestimate the total free energy of the
active region by a factor of $\approx 4$ in this case. In the other 
two cases we analyzed, which are further away from the solar disk center 
and provide less accurate measurements of magnetic energies, 
we find that AIA underestimates the free energy derived from ROSA 
by factors of 1.5-2.5, and thus reveal the same trend.}

\item{\underbar{Contribution function of altitude:}
The 3-D magnetic field model of the VCA-NLFFF code yields also
a fit of the line-of-sight coordinate $z(s)$ to each curvi-linear loop 
segment that is traced in the $[x,y]$ plane with the OCCULT-2 code.
From the so obtained 3-D coordinates $[x,y,z]$ we can 
immediately derive the altitude $h$ of
each traced loop segment. Plotting histograms of these altitudes we find that the
IBIS, ROSA, and IRIS data all reveal chromospheric and transition region
structures in a height range of $h \lapprox 4000$ km, which is
consistent with the height range of fibrils, filaments, mottles, and 
spicules in the dynamic part of the chromosphere, which extend beyond the
hydrostatic height range of $h \lapprox 2000$ km. In contrast, the
AIA 171, 193, 211 \ang , as well as the AIA 304 \ang\ wavelength 
in the cooler He II line reveal cool plasma structures 
in a height range of $h \lapprox 35,000$ km.
Apparently, the features seen in He II include mostly cool plasma
at coronal heights, including ``coronal rain'', filaments, and prominences,
rather than structures in the chromosphere.}

\item{\underbar{Plasma $\beta$-parameter:}
From our 3-D magnetic field model obtained with VCA-NLFFF we can
also directly calculate the magnetic field $B(h)$ as a function of height
for each traced structure, as well as the plasma-$\beta$ parameter
$\beta(h)$ as a function of height. Interestingly we find that all traced
structures have magnetic fields in the range of $B(h) \approx 100-1000$ G,
and a plasma-$\beta$ parameter in the range of $\beta(h) \approx 10^{-5}-10^{-1}$,
over the entire mapped height range of $h \approx 100-35,000$ km. Thus all
traced structures are magnetically confined, which can be explained by their
proximity to the dominant sunspot. However, it is surprising that our automated
feature detection code did not pick up any structure (out of the $\approx
1500$ field-aligned structures found in all fits) in regions with a low
magnetic field, and possibly with a plasma-$\beta$ parameter in excess of unity.
This finding has perhaps the important consequence that coronal heating
occurs mostly in strong-field regions with $B \gapprox 100$ G, rather than in
weak-field regions.}

\end{enumerate}

In summary, our study has shown that high-resolution chromospheric images
are extremely useful for magnetic modeling, equally important as 
(high-resolution) coronal images. Obviously, the optimum wavelengths are
in the core of UV and H-$\alpha$ line profiles, which have a
peak of the height contribution function in the upper chromosphere.
The features that constrain magnetic field models best are crisp
curvi-linear structures, such as loops, fibrils, filaments, spicules,
and threads of prominences. We learned that chromospheric features may
even yield more comprehensive estimates of the free magnetic energy than
coronal loops, and thus both data sets should be combined for magnetic
modeling in future efforts. The features seen in chromospheric images
complement those seen in coronal images, which probe two different
but complementary height ranges. The reliability of any (non-potential) 
magnetic field solution obtained with either chromospheric or coronal 
images depends strongly on the selection of traced features, which should 
ideally comprise a representative subset of all magnetic field structures 
in a given computation box of an active region, sampled in both the
chromospheric and coronal height range. All these conclusions strongly
suggest that chromospheric and coronal data need to be quantitatively
included in magnetic field models of solar flares and active regions, 
a requirement that is incorporated in the present VCA-NLFFF method, 
or in the quasi-Grad-Rubin method of Malanushenko et al.~(2014), 
while there is no provision for such a capability in traditional 
NLFFF codes. 

\bigskip
\acknowledgements
The author is indebted to helpful discussions with Bart DePontieu,
Mark DeRosa, Anna Malanushenko, Carolus Schrijver, Alberto Sainz-Dalda, 
Ada Ortiz, and Jorrit Leenaarts. Part of the work was supported by the 
NASA contracts NNG04EA00C of the SDO/AIA instrument and 
NNG09FA40C of the IRIS mission.
D.B.J. thanks the UK's Science and Technology Facilities Council (STFC) 
for an Ernest Rutherford Fellowship, in addition to a dedicated standard 
grant, which allowed this project to be undertaken.

\clearpage



\begin{deluxetable}{lll}
\tabletypesize{\normalsize}
\tablecaption{Data selection parameters and adjustable control parameters
of the VCA-NLFFF forward-fitting code used in this study.}
\tablewidth{0pt}
\tablehead{
\colhead{Task:}&
\colhead{Control parameter}&
\colhead{Value}}
\startdata
Data selection: &Instruments                    & HMI; AIA; IRIS; IBIS; ROSA                \\
                &Spatial pixel size             & $0.5\arcsec$; $0.6\arcsec$;
                                                  $0.16\arcsec$; $0.1\arcsec$; $0.1\arcsec$ \\
                &Wavelengths                    & 6173; [94,131,171,193,211,304,335,1600];  \\
                &                               & [1400,2796,2832]; 8542; 6563 \ang         \\
                &Field-of-view                  & FOV = 0.1 $R_{\odot}$                     \\
Magnetic sources: &Number of magnetic sources   & $n_{mag} = 30, (50)$                      \\
                &Width of fitted local maps     & $w_{mag}=3$ pixels                        \\
                &Depth range of buried charges  & $d_{mag} = 20$ pixels                     \\
                &Rebinned pixel size            & $\Delta x_{mag}=3$ pixels = $1.5\arcsec$  \\
Loop tracing:   &Maximum of traced structures   & $n_{struc}=1000$                          \\
                &Lowpass filter                 & $n_{sm1} = 1$, (3) pixel                  \\
                &Highpass filter                & $n_{sm2} = n_{sm1}+2 = 3$ pixels          \\
                &Minimum loop length            & $l_{min} = 5$, (4) pixels                 \\
                &Minimum loop curvature radius  & $r_{min} = 8$ pixels                      \\
                &Field line step                & $\Delta s=0.002 R_{\odot}$                \\
                &Threshold positive flux        & $q_{thresh,1} = 0$                        \\
                &Threshold positive filter flux & $q_{thresh,2} = 0$                        \\
                &Proximity to magnetic sources  & $d_{prox}=10$ source depths               \\
Forward-Fitting:&Misalignment angle limit       & $\mu_0 = 20^\circ, (30^\circ)$            \\
                &Minimum number of iterations   & $n_{iter,min}= 40$                        \\
                &Maximum number of iterations   & $n_{iter,max}= 100$                       \\
                &Number loop segment positions  & $n_{seg}=9$                               \\
                &Maximum altitude               & $h_{max}=0.2 R_{\odot}$                   \\
                &$\alpha$-parameter increment   & $\Delta \alpha_0=1.0 \ R_{\odot}^{-1}$    \\
                &Isotropic current correction   & $q_{iso}=(\pi/2)^2\approx 2.5$            \\
\enddata
\end{deluxetable}

\begin{deluxetable}{llllll}
\tabletypesize{\normalsize}
\tablecaption{Observation date and times, active region numbers, instruments,
and wavelength ranges.}
\tablehead{
\colhead{Observation}&
\colhead{Active}&
\colhead{Heliographic}&
\colhead{Instrument}&
\colhead{Wavelength}&
\colhead{Temperature}\\
\colhead{date and time}&
\colhead{Region}&
\colhead{Position}&
\colhead{}&
\colhead{}&
\colhead{range}\\
\colhead{[UT]}&
\colhead{NOAA}&
\colhead{[deg]}&
\colhead{}&
\colhead{\ang}&
\colhead{log(T[K])}}
\startdata
      2010-08-03 15:23:00 &11092 &N12W02 &AIA  & 171, 193, 211 & 5.8-7.3 \\         
      2010-08-03 15:23:00 &11092 &N12W02 &AIA  & 304           & 4.7     \\ 
      2010-08-03 15:23:00 &11092 &N12W02 &IBIS &8542           & 3.8     \\ 
      2010-08-03 15:23:00 &11092 &N12W02 &IBIS &6563           & 3.8     \\        
                          &      &       &     &               & \\
      2014-08-24 14:06:38 &12146 &N09W25 &AIA  & 171, 193, 211 & 5.8-7.3 \\           
      2014-08-24 14:06:38 &12146 &N09W25 &AIA  & 304           & 4.7     \\         
      2014-08-24 14:06:38 &12146 &N09W25 &ROSA &6563           & 3.8     \\         
                          &      &       &     &               & \\
      2014-08-30 14:40:22 &12149 &N12W44 &AIA  & 171, 193, 211 & 5.8-7.3 \\          
      2014-08-30 14:40:22 &12149 &N12W44 &AIA  & 304           & 4.7     \\        
      2014-08-30 14:40:22 &12149 &N12W44 &IRIS &2796           & 3.7-4.2 \\        
      2014-08-30 14:40:22 &12149 &N12W44 &ROSA &6563           & 3.8     \\         
\enddata
\end{deluxetable}

\begin{deluxetable}{llrrrrcr}
\tabletypesize{\normalsize}
\tablecaption{Data analysis results of the number of detected (field-aligned)
loops $n_{det}$, fitted loops $n_{loop}$, the 2-D and 3-D misalignment angles 
$\mu_2$ and $\mu_3$, the potential energy $E_P$, and the ratio of the free energy
$q_{free}=E_{NP}/E_P-1$. The means and error bars are averaged from fitting
5 different variations of the control parameters.}
\tablehead{
\colhead{Observation}&
\colhead{Instrument}&
\colhead{Detected}&
\colhead{Fitted}&
\colhead{Misalignment}&
\colhead{Misalignment}&
\colhead{Potential}&
\colhead{Free energy}\\
\colhead{date}&
\colhead{}&
\colhead{loops}&
\colhead{loops}&
\colhead{angle 2-D}&
\colhead{angle 3-D}&
\colhead{energy}&
\colhead{ratio}\\
\colhead{}&
\colhead{}&
\colhead{$n_{det}$}&
\colhead{$n_{loop}$}&
\colhead{$\mu_2$ [deg]}&
\colhead{$\mu_3$ [deg]}&
\colhead{$E_P$ [$10^{30}$ erg]}&
\colhead{$q_{free}$}}
\startdata
2010-08-03 & AIA 171+  & $ 222\pm 76$ &  $ 167\pm 36$ &  $   5.0^\circ\pm 3.2^\circ$ &  $   7.8^\circ\pm 0.5^\circ$ &  $   571$ &  $  0.03\pm0.01$ \\
2010-08-03 & AIA 304   & $  63\pm 27$ &  $  38\pm 14$ &  $   4.2^\circ\pm 0.4^\circ$ &  $   7.2^\circ\pm 0.9^\circ$ &  $   571$ &  $  0.06\pm0.03$ \\
2010-08-03 & IBIS 8542 & $ 656\pm121$ &  $ 338\pm 62$ &  $   4.0^\circ\pm 0.6^\circ$ &  $   7.1^\circ\pm 0.7^\circ$ &  $   571$ &  $  0.13\pm0.04$ \\
2010-08-03 & IBIS 6563 & $ 712\pm114$ &  $ 421\pm 75$ &  $   4.0^\circ\pm 0.4^\circ$ &  $   7.2^\circ\pm 0.8^\circ$ &  $   571$ &  $  0.11\pm0.01$ \\
           &           &              &               &                              &                              &           &                  \\
2014-08-24 & AIA 171+  & $ 186\pm 88$ &  $  82\pm 30$ &  $   5.5^\circ\pm 1.3^\circ$ &  $   8.9^\circ\pm 1.5^\circ$ &  $   551$ &  $  0.18\pm0.06$ \\
2014-08-24 & AIA 304   & $  45\pm 21$ &  $  17\pm  6$ &  $   7.4^\circ\pm 2.2^\circ$ &  $   7.5^\circ\pm 1.5^\circ$ &  $   551$ &  $  0.11\pm0.05$ \\
2014-08-24 & ROSA 6563 & $ 654\pm 98$ &  $ 232\pm 75$ &  $   6.5^\circ\pm 1.3^\circ$ &  $   8.5^\circ\pm 1.8^\circ$ &  $   551$ &  $  0.26\pm0.01$ \\
           &           &              &               &                              &                              &           &                  \\
2014-08-30 & AIA 171+  & $ 190\pm 87$ &  $  83\pm 34$ &  $   5.4^\circ\pm 1.2^\circ$ &  $  10.9^\circ\pm 2.4^\circ$ &  $   559$ &  $  0.10\pm0.05$ \\
2014-08-30 & AIA 304   & $  43\pm 22$ &  $  16\pm  7$ &  $   7.0^\circ\pm 1.0^\circ$ &  $  10.7^\circ\pm 2.4^\circ$ &  $   559$ &  $  0.10\pm0.09$ \\
2014-08-30 & IRIS 2796 & $ 206\pm 52$ &  $  65\pm 19$ &  $   6.1^\circ\pm 1.4^\circ$ &  $  11.3^\circ\pm 2.5^\circ$ &  $   559$ &  $  0.26\pm0.01$ \\
2014-08-30 & ROSA 6563 & $ 556\pm 89$ &  $ 299\pm 79$ &  $   6.6^\circ\pm 0.5^\circ$ &  $  14.2^\circ\pm 2.3^\circ$ &  $   559$ &  $  0.17\pm0.04$ \\
\enddata
\end{deluxetable}

\clearpage

\begin{figure}
\plotone{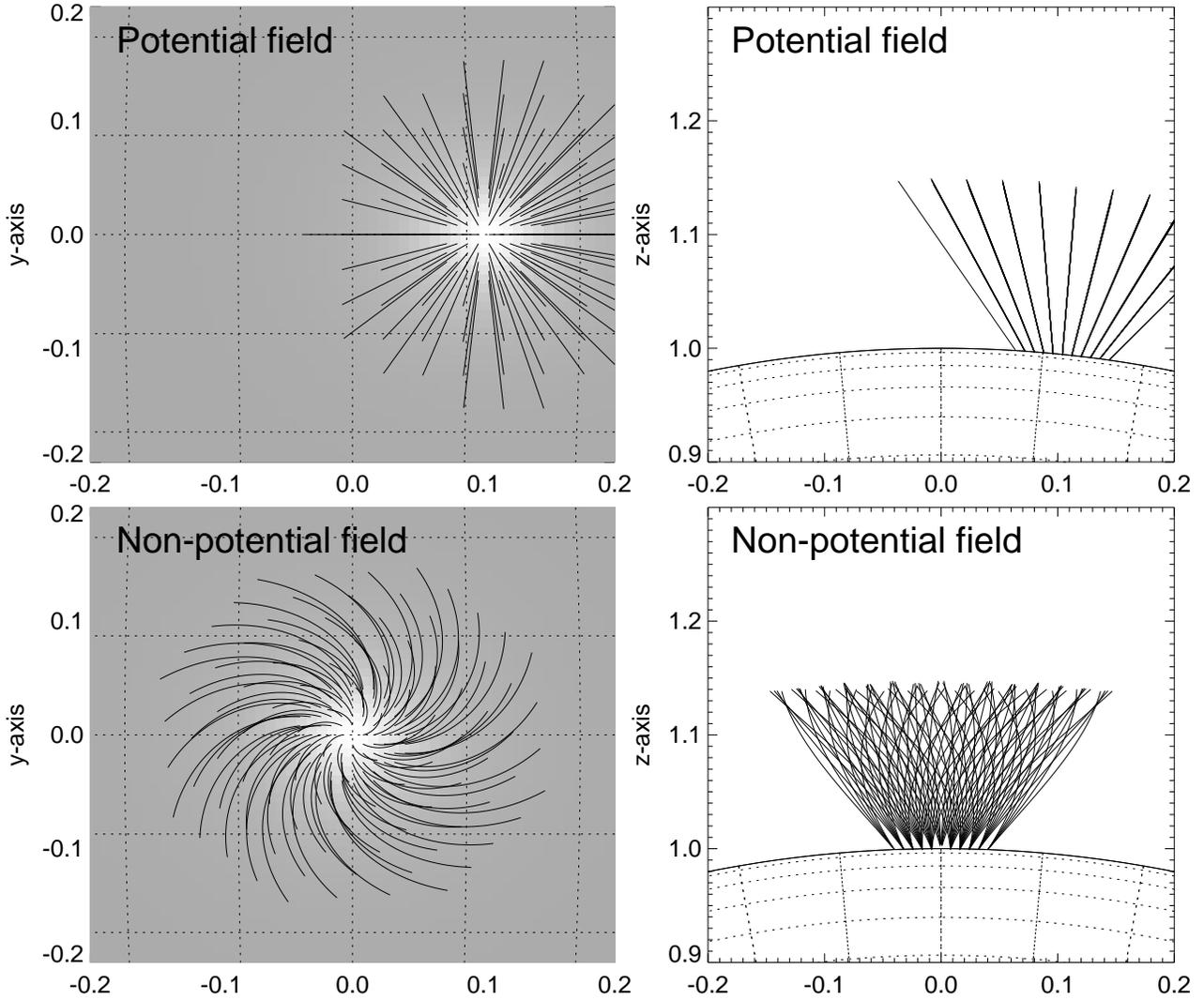}
\caption{Magnetic field lines of a single unipolar magnetic charge, 
mimicking 
a single sunspot, computed for a potential field (top panels) and for a 
non-potential field based on the vertical-current approximation
(VCA-NLFFF code), which 
introduces a helical twist about the vertical axis (bottom panels). 
The left panels depict a projection from top down to the solar surface, 
while the right panels show a side view. The greyscale indicates the
corresponding line-of-sight magnetogram.} 
\end{figure}

\begin{figure}
\plotone{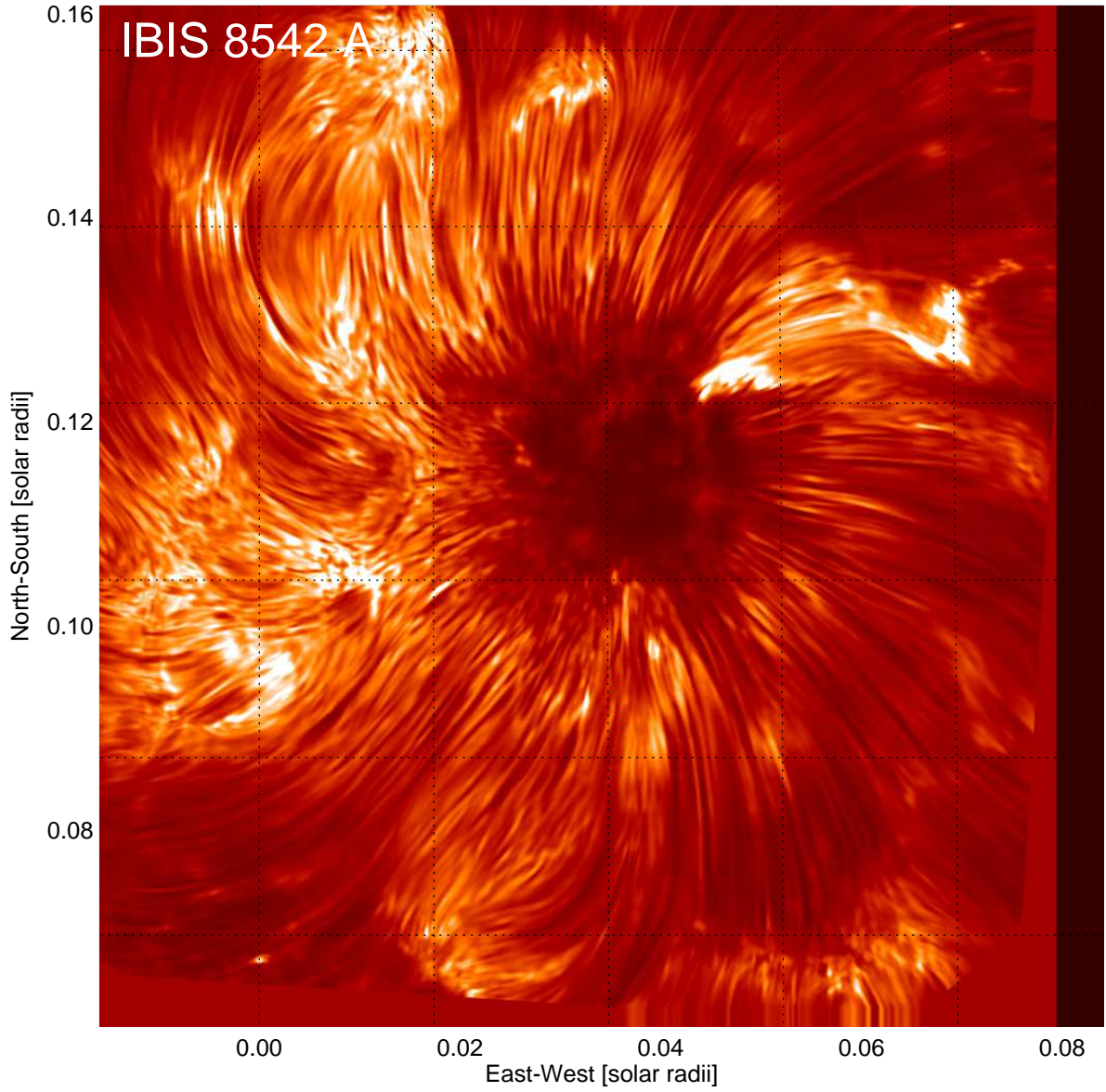}
\caption{IBIS image of active region NOAA 11092 observed on 2010 Aug 3, 
15:03-15:43 UT
in the Ca II 8542 \ang\ line.  The displayed field-of-view captures a 
subimage centered at the sunspot with a width of FOV = 0.10 $R_{\odot}$ and 
a spatial pixel size of $0.1\arcsec$.}
\end{figure}

\begin{figure}
\plotone{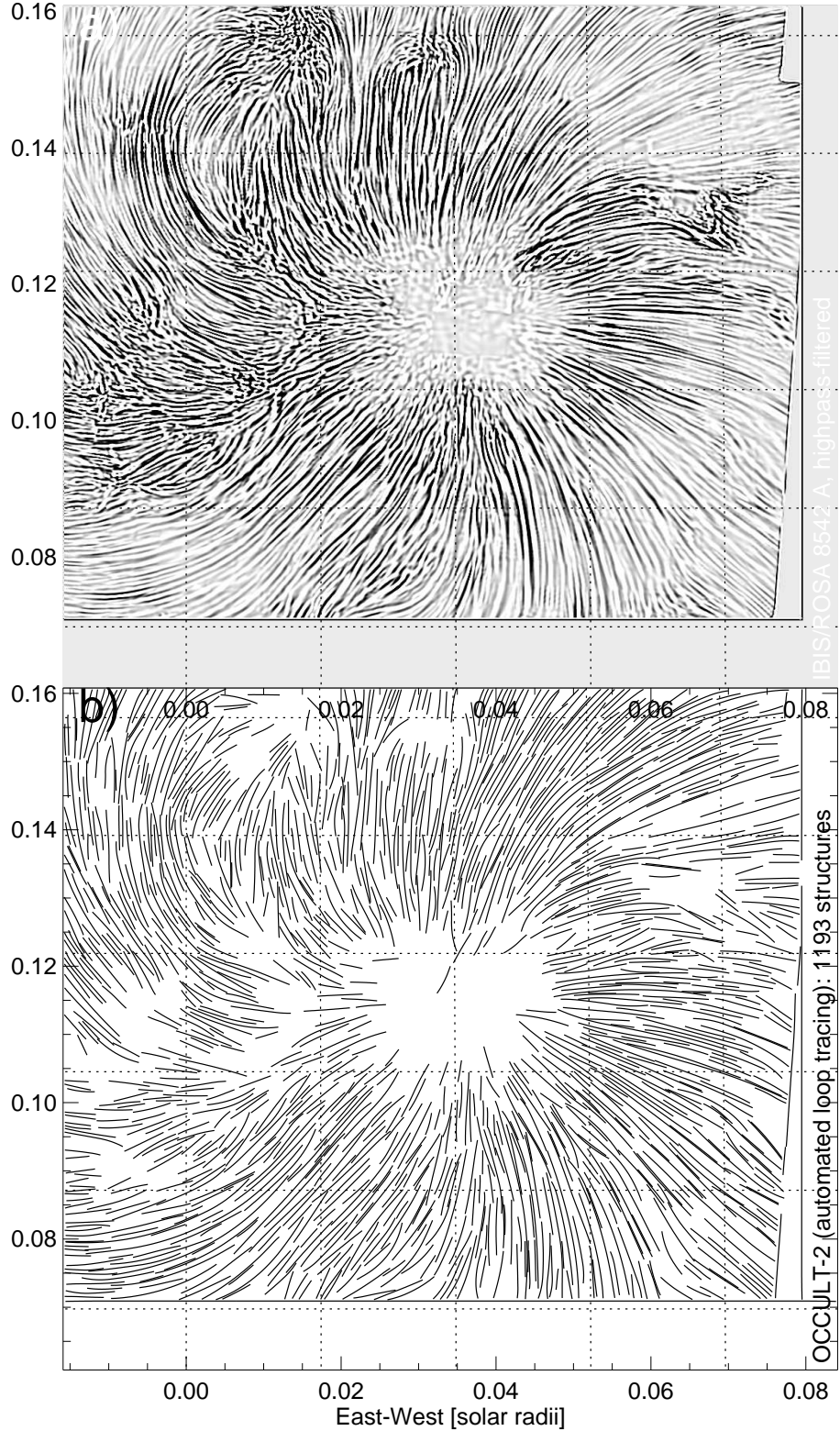}
\caption{A highpass-filtered version of the original IBIS 8542 \ang\
subimage shown in Fig.~2 (top panel), along with 1193 automatically 
traced curvi-linear structures using the OCCULT-2 code (bottom panel), 
which may consist of chromospheric fibrils or footpoints of coronal loops.} 
\end{figure}

\begin{figure}
\plotone{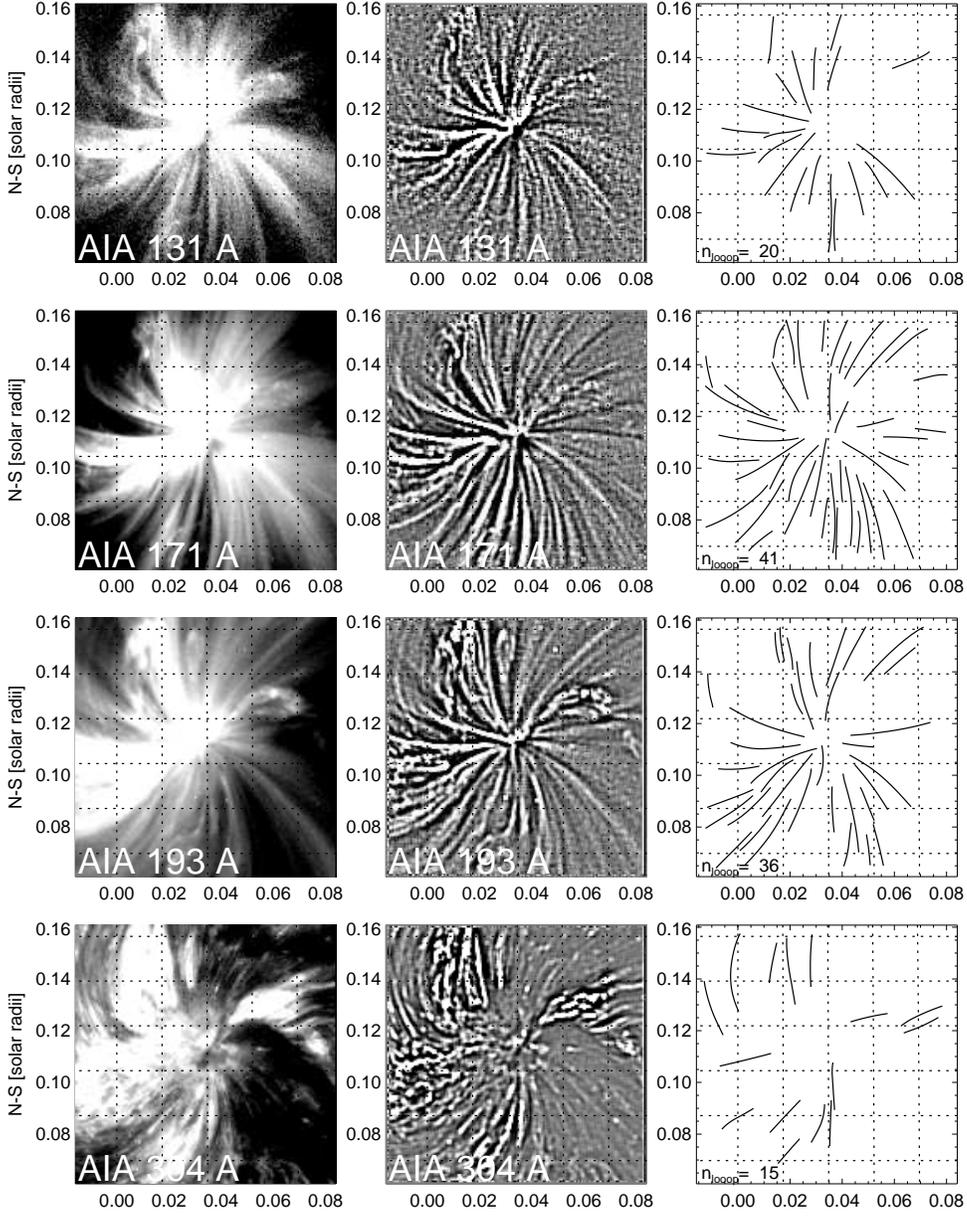}
\caption{AIA/SDO images in the wavelengths of 131, 171, 193, and 304
\ang\ on a logarithmic scale (left column), their highpass-filtered 
counterparts on a linear scale and smoothed with a boxcar of $n_{sm1}=3$
pixels (middle column), and the results of automated loop tracings 
obtained with the OCCULT-2 code (right column).}
\end{figure}

\begin{figure}
\plotone{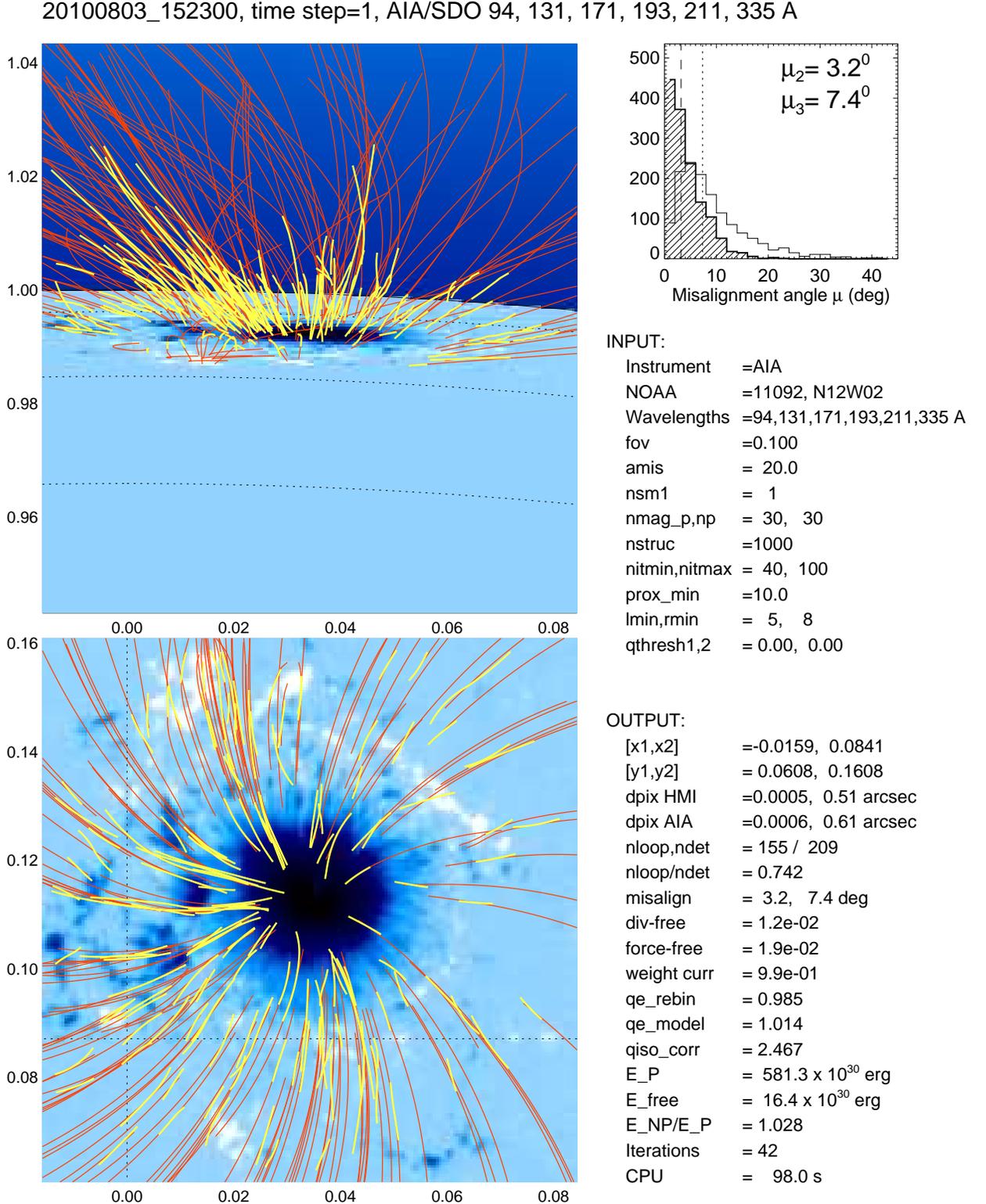}
\caption{The automated curvi-linear feature tracing in the AIA images 
(2010 Aug 03, 15:23 UT) in
6 AIA wavelengths of 94, 131, 171, 193, 211, 335 \ang\ (yellow curves) 
are shown, 
overlaid on the best-fit solutions of the magnetic field model using the 
VCA-NLFFF code (red curves), and the observed HMI magnetogram (blue 
background image), from the line-of-sight view in the $(x,y)$-plane 
(bottom panel) and the orthogonal projection in the $(x,z)$-plane (top panel). 
A histogram of the 2-D and 3-D misalignment angles and various input and 
output parameters are shown in the top right-hand panel.}
\end{figure}

\begin{figure}
\plotone{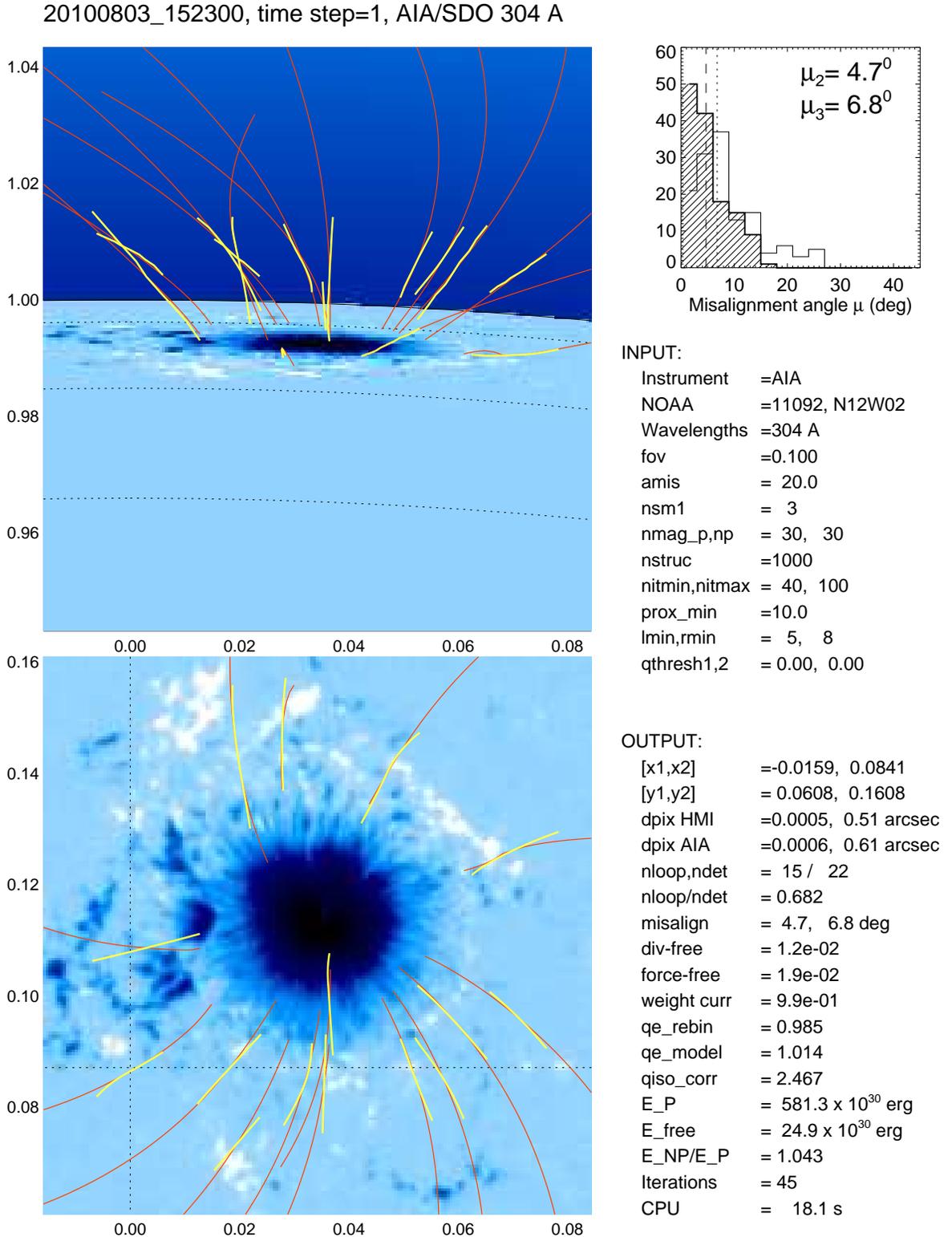}
\caption{Automated feature tracking is applied to the AIA  
image (2010 Aug 03, 15:23 UT) in the wavelength of He I 304 \ang\ wavelength,
otherwise similar representation as in Fig.~5.}
\end{figure}

\begin{figure}
\plotone{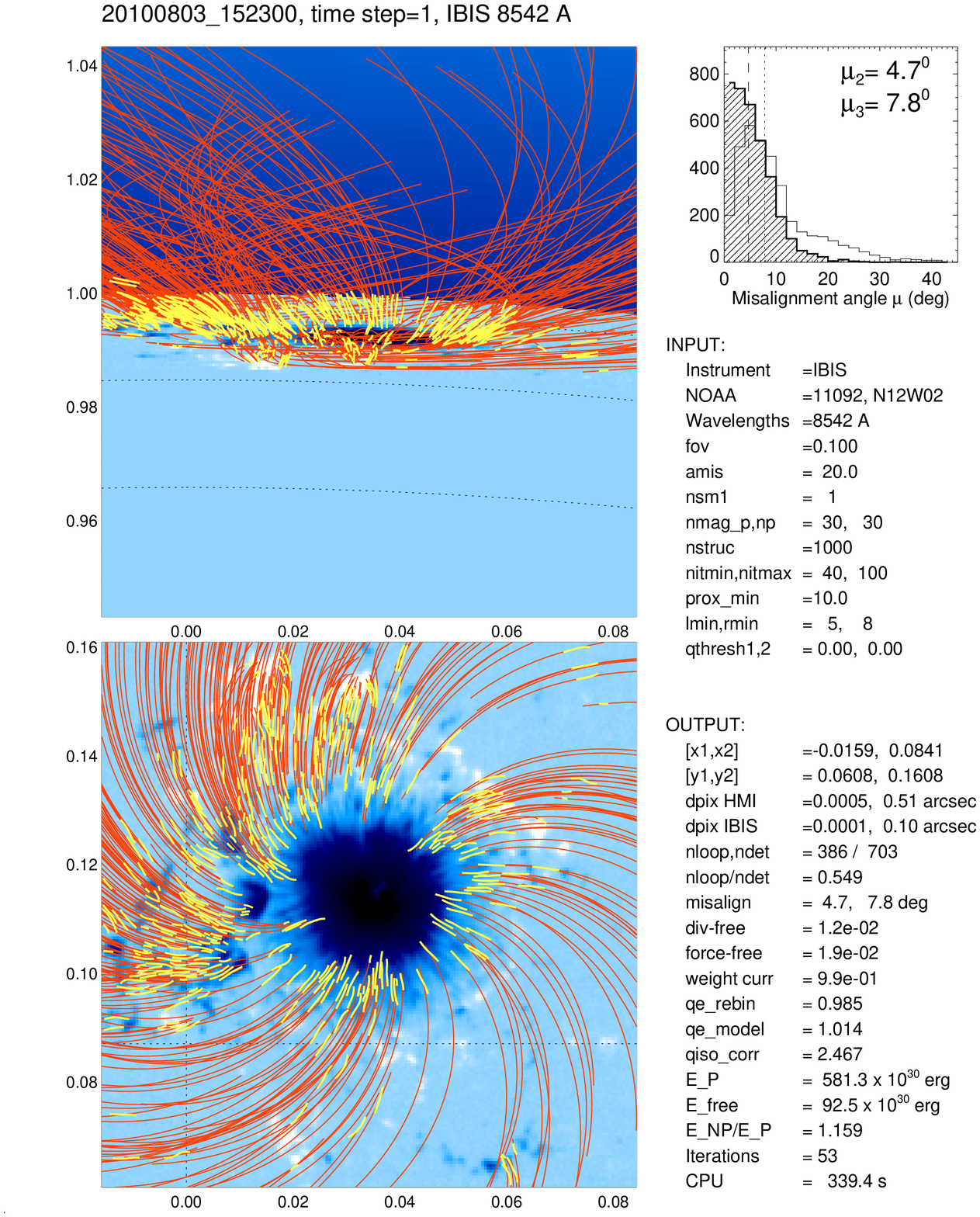}
\caption{Automated feature tracking is applied to the chromospheric IBIS 
image (2010 Aug 03, 15:23 UT) in the wavelength of Ca II 8542 \ang\ wavelength,
otherwise similar representation as in Fig.~5.}
\end{figure}

\begin{figure}
\plotone{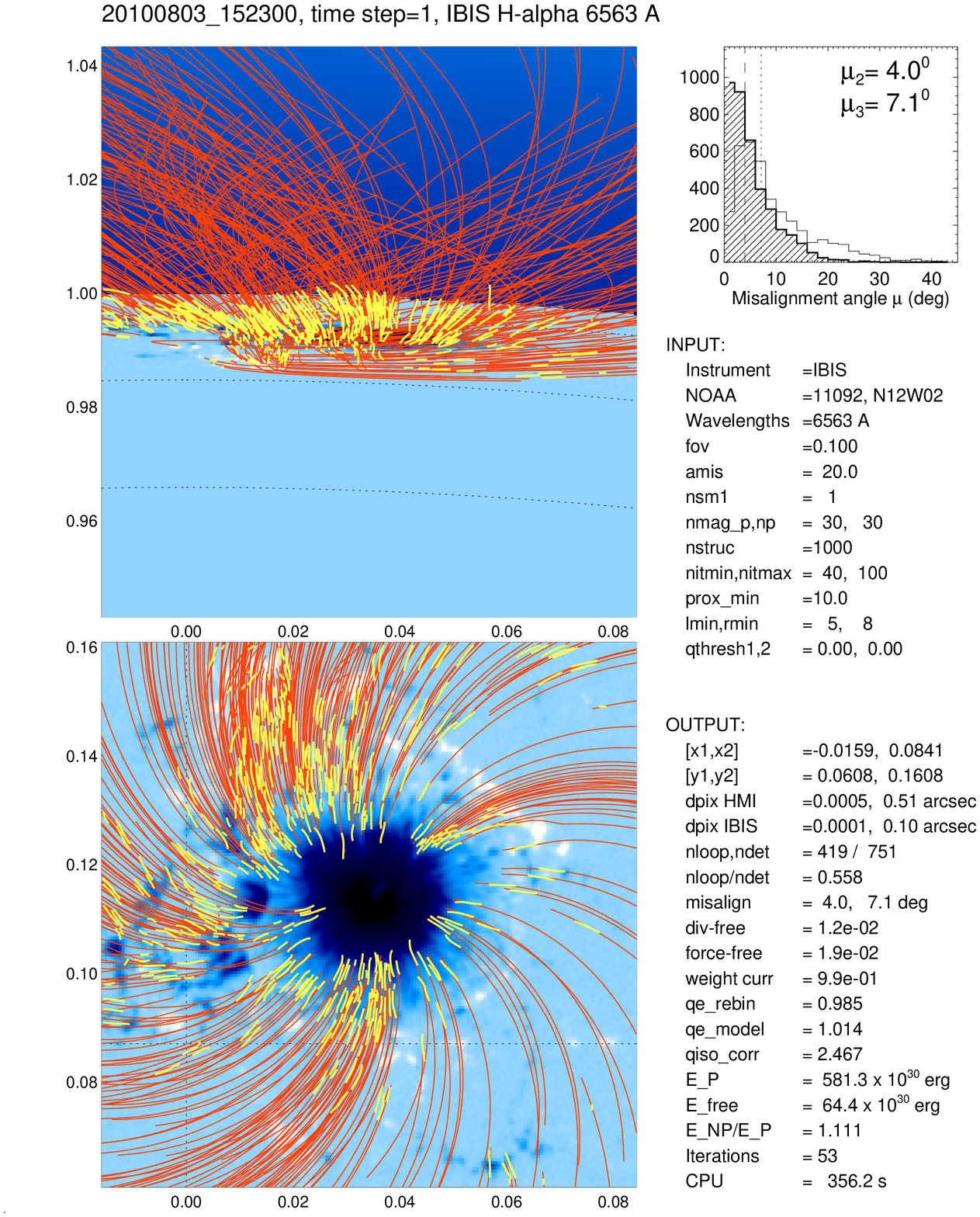}
\caption{Automated feature tracking is applied to the chromospheric IBIS 
image (2010 Aug 03, 15:23 UT) in the wavelength of H$\alpha$ 6563 \ang\ wavelength,
otherwise similar representation as in Fig.~5.}
\end{figure}

\begin{figure}
\plotone{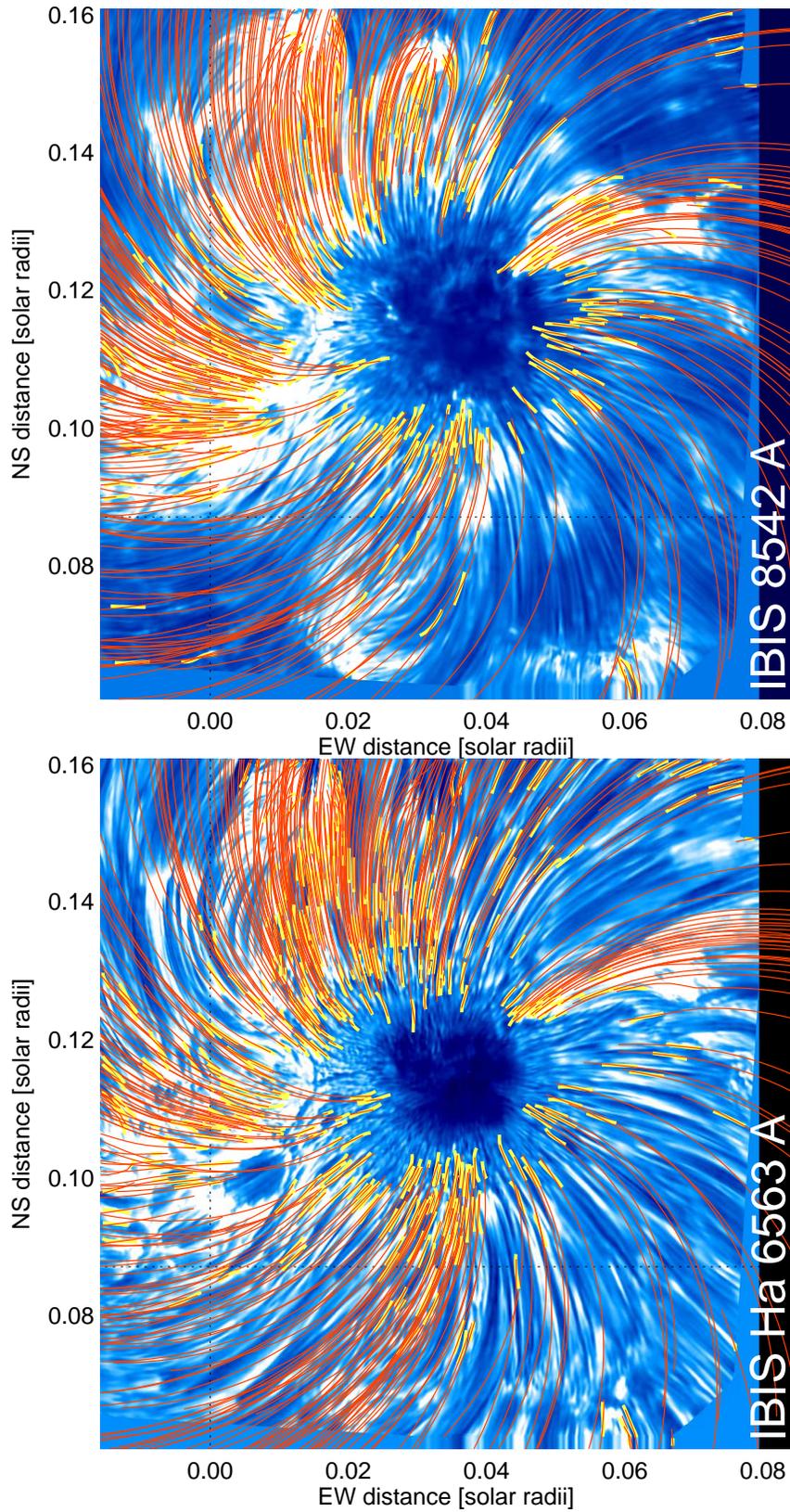}
\caption{A subset of automatically traced loop segments (yellow
curves) and best-fit magnetic field lines (red curves) are shown,
overlaid on the IBIS 8542 \ang\ image (top frame) and 
H$\alpha$ 6563 \ang\ image (bottom frame) in which the
automated tracing was performed.}
\end{figure}

\begin{figure}
\plotone{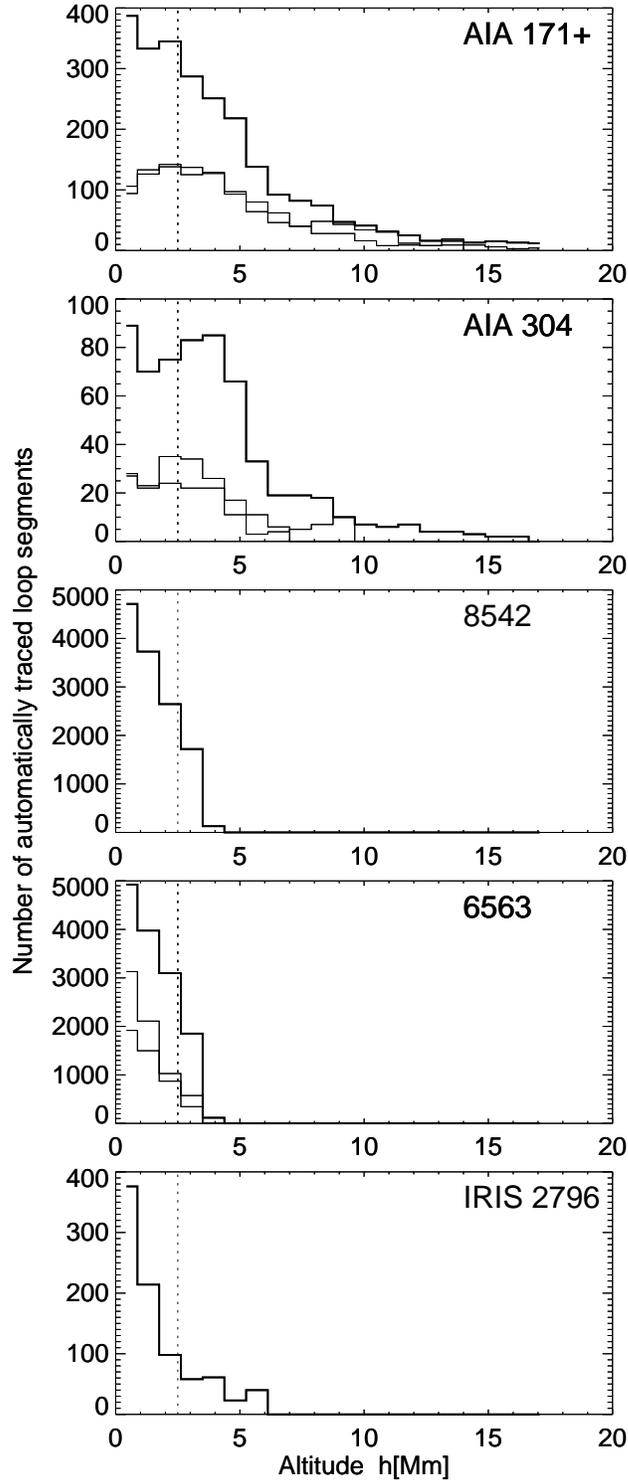}
\caption{Distribution function of altitudes $h$ of automatically traced
curvi-linear elements according to the 3D modeling of the forward-fitting 
VCA-NLFFF code. Most of the 
features seen with IBIS, ROSA, and IRIS originate in the chromosphere 
(marked at a nominal height of 2500 km with a dotted line) 
and transition region, while most of the structures seen with AIA belong to the
corona, including the emission at 304 \ang . The panels show one, two, or three
histograms, depending on the number of observing days as listed in Table 2.
The highest histogram always refers to the observations of NOAA 11092 on
2010 Aug 3, which is also closest to disk center and exhibits the largest
number of loops.}
\end{figure}

\begin{figure}
\plotone{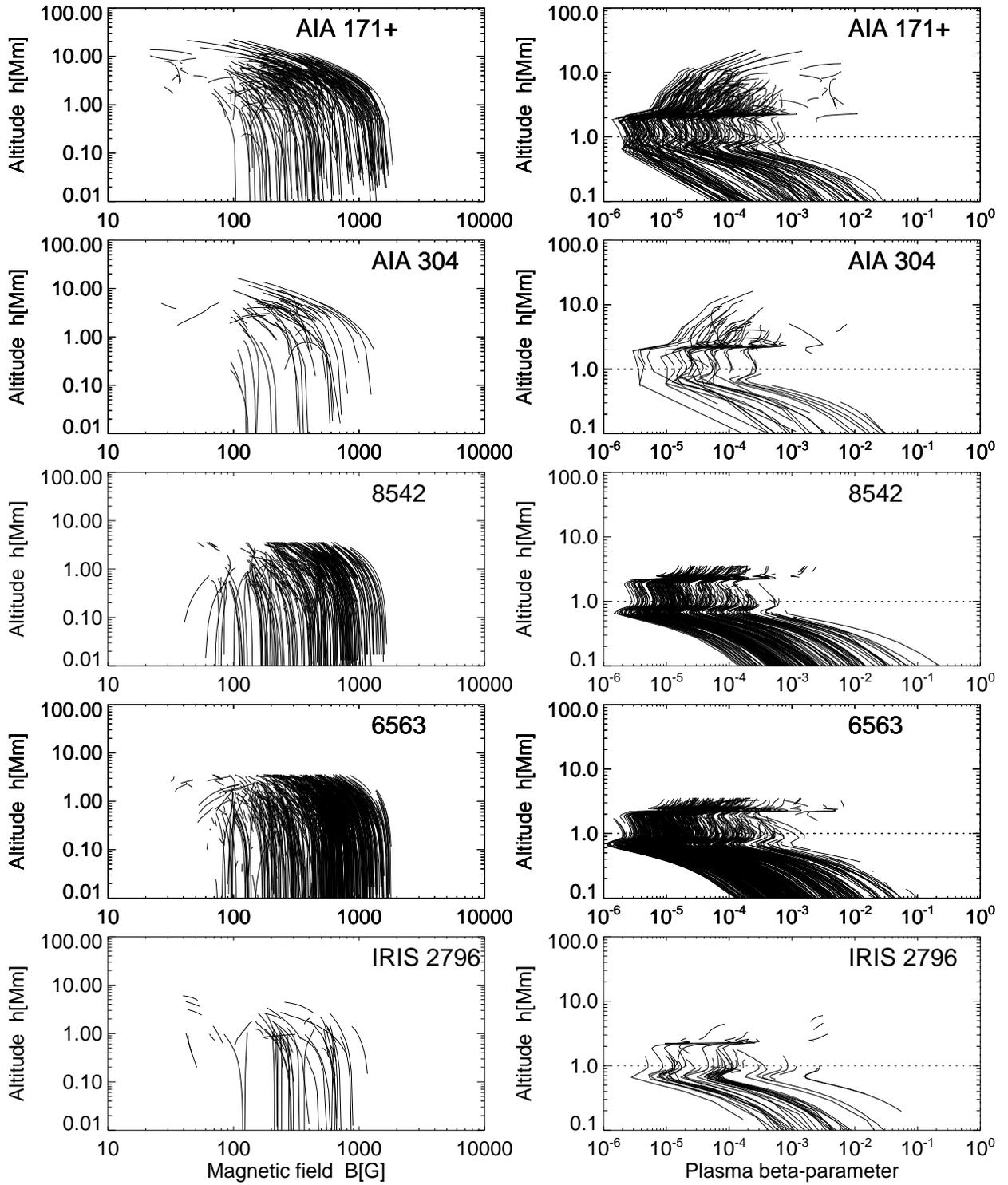}
\caption{The magnetic field strength $B(h)$ as a function of the height $h$
(left panels) and the plasma-$\beta$ parameter $\beta(h)$ as a function of
height $h$ (right panels), calculated from the chromospheric VAL-C model
(Fontenla et al.~1990) and the coronal model by Gabriel (1976), for all
automatically traced curvi-linear structures observed with each instrument
(AIA, IRIS, IBIS, ROSA).}
\end{figure}

\end{document}